\PassOptionsToPackage{unicode}{hyperref}
\PassOptionsToPackage{hyphens}{url}
\PassOptionsToPackage{dvipsnames,svgnames,x11names}{xcolor}
\documentclass[
]{article}
\usepackage{xcolor}
\usepackage{amsmath,amssymb}
\usepackage{iftex}
\ifPDFTeX
  \usepackage[T1]{fontenc}
  \usepackage[utf8]{inputenc}
  \usepackage{textcomp} 
\else 
  \usepackage{unicode-math} 
  \defaultfontfeatures{Scale=MatchLowercase}
  \defaultfontfeatures[\rmfamily]{Ligatures=TeX,Scale=1}
\fi
\usepackage{lmodern}
\ifPDFTeX\else
\fi
\IfFileExists{upquote.sty}{\usepackage{upquote}}{}
\IfFileExists{microtype.sty}{
  \usepackage[]{microtype}
  \UseMicrotypeSet[protrusion]{basicmath} 
}{}
\makeatletter
\@ifundefined{KOMAClassName}{
  \IfFileExists{parskip.sty}{%
    \usepackage{parskip}
  }{
    \setlength{\parindent}{0pt}
    \setlength{\parskip}{6pt plus 2pt minus 1pt}}
}{
  \KOMAoptions{parskip=half}}
\makeatother
\makeatletter
\ifx\paragraph\undefined\else
  \let\oldparagraph\paragraph
  \renewcommand{\paragraph}{
    \@ifstar
      \xxxParagraphStar
      \xxxParagraphNoStar
  }
  \newcommand{\xxxParagraphStar}[1]{\oldparagraph*{#1}\mbox{}}
  \newcommand{\xxxParagraphNoStar}[1]{\oldparagraph{#1}\mbox{}}
\fi
\ifx\subparagraph\undefined\else
  \let\oldsubparagraph\subparagraph
  \renewcommand{\subparagraph}{
    \@ifstar
      \xxxSubParagraphStar
      \xxxSubParagraphNoStar
  }
  \newcommand{\xxxSubParagraphStar}[1]{\oldsubparagraph*{#1}\mbox{}}
  \newcommand{\xxxSubParagraphNoStar}[1]{\oldsubparagraph{#1}\mbox{}}
\fi
\makeatother

\usepackage{longtable,booktabs,array}
\usepackage{calc} 
\usepackage{etoolbox}
\makeatletter
\patchcmd\longtable{\par}{\if@noskipsec\mbox{}\fi\par}{}{}
\makeatother
\IfFileExists{footnotehyper.sty}{\usepackage{footnotehyper}}{\usepackage{footnote}}
\makesavenoteenv{longtable}
\usepackage{graphicx}
\makeatletter
\newsavebox\pandoc@box
\newcommand*\pandocbounded[1]{
  \sbox\pandoc@box{#1}%
  \Gscale@div\@tempa{\textheight}{\dimexpr\ht\pandoc@box+\dp\pandoc@box\relax}%
  \Gscale@div\@tempb{\linewidth}{\wd\pandoc@box}%
  \ifdim\@tempb\p@<\@tempa\p@\let\@tempa\@tempb\fi
  \ifdim\@tempa\p@<\p@\scalebox{\@tempa}{\usebox\pandoc@box}%
  \else\usebox{\pandoc@box}%
  \fi%
}
\def\fps@figure{htbp}
\makeatother

\NewDocumentCommand\citeproctext{}{}

\makeatletter
 \let\@cite@ofmt\@firstofone
 \def\@biblabel#1{}
 \def\@cite#1#2{{#1\if@tempswa , #2\fi}}
\makeatother
\newlength{\cslhangindent}
\newlength{\csllabelwidth}
\newenvironment{CSLReferences}[2] 
 {\begin{list}{}{%
  \setlength{\itemindent}{0pt}
  \setlength{\leftmargin}{0pt}
  \setlength{\parsep}{0pt}
  \ifodd #1
   \setlength{\leftmargin}{\cslhangindent}
   \setlength{\itemindent}{-1\cslhangindent}
  \fi
  \setlength{\itemsep}{#2\baselineskip}}}
 {\end{list}}
\usepackage{calc}

\usepackage{arxiv}
\usepackage{orcidlink}
\usepackage{amsmath}
\usepackage[T1]{fontenc}
\makeatletter
\@ifpackageloaded{caption}{}{\usepackage{caption}}
\AtBeginDocument{%
\ifdefined\contentsname
  \renewcommand*\contentsname{Table of contents}
\else
  \newcommand\contentsname{Table of contents}
\fi
\ifdefined\listfigurename
  \renewcommand*\listfigurename{List of Figures}
\else
  \newcommand\listfigurename{List of Figures}
\fi
\ifdefined\listtablename
  \renewcommand*\listtablename{List of Tables}
\else
  \newcommand\listtablename{List of Tables}
\fi
\ifdefined\figurename
  \renewcommand*\figurename{Figure}
\else
  \newcommand\figurename{Figure}
\fi
\ifdefined\tablename
  \renewcommand*\tablename{Table}
\else
  \newcommand\tablename{Table}
\fi
}
\@ifpackageloaded{float}{}{\usepackage{float}}
\floatstyle{ruled}
\@ifundefined{c@chapter}{\newfloat{codelisting}{h}{lop}}{\newfloat{codelisting}{h}{lop}[chapter]}
\floatname{codelisting}{Listing}

\makeatother
\makeatletter
\@ifpackageloaded{caption}{}{\usepackage{caption}}
\@ifpackageloaded{subcaption}{}{\usepackage{subcaption}}
\makeatother
\usepackage{bookmark}
\IfFileExists{xurl.sty}{\usepackage{xurl}}{} 
\hypersetup{
  pdftitle={Exploring Students' Understanding of Linear and Quadratic Relationships in a Projectile Motion Context},
  pdfauthor={Yosep Dwi Kristanto; Teo Paoletti; Russasmita Sri Padmi; Serli Evidiasari; Zsolt Lavicza; Tony Houghton; Houssam Kasti},
  pdfkeywords={covariational reasoning, linear relationship, projectile
motion, quadratic relationship},
  colorlinks=true,
  linkcolor={blue},
  filecolor={Maroon},
  citecolor={Blue},
  urlcolor={Blue},
  pdfcreator={LaTeX via pandoc}}

\newcommand{\runninghead}{A Preprint }
\renewcommand{\runninghead}{Author Accepted Manuscript }
\title{Exploring Students' Understanding of Linear and Quadratic
Relationships in a Projectile Motion Context}
\def\asep{\\\\\\ } 
\def\asep{\And }
\author{\textbf{Yosep Dwi
Kristanto}~\orcidlink{0000-0003-1446-0422}\\Department for STEM
Didactics\\Johannes Kepler University Linz\\Linz,\ 4040\\Mathematics
Education Study Program\\Sanata Dharma
University\\Yogyakarta\\\href{mailto:yosepdwikristanto@usd.ac.id}{yosepdwikristanto@usd.ac.id}\asep\textbf{Teo
Paoletti}~\orcidlink{0000-0002-1377-1133}\\School of
Education\\University of Delaware\\Newark, DE\\\asep\textbf{Russasmita
Sri Padmi}~\orcidlink{0000-0001-7180-459X}\\Department for STEM
Didactics\\Johannes Kepler University
Linz\\Linz,\ 4040\\\asep\textbf{Serli Evidiasari}\\\\SMP Negeri 4
Pakem\\Sleman\\\asep\textbf{Zsolt
Lavicza}~\orcidlink{0000-0002-3701-5068}\\Department for STEM
Didactics\\Johannes Kepler University
Linz\\Linz,\ 4040\\\asep\textbf{Tony
Houghton}~\orcidlink{0000-0002-2899-3310}\\Department for STEM
Didactics\\Johannes Kepler University
Linz\\Linz,\ 4040\\\asep\textbf{Houssam
Kasti}~\orcidlink{0000-0001-6573-8370}\\Faculty of General
Education\\Qatar University\\Doha\\}
\date{}
\begin{document}
\maketitle
\begin{abstract}
Previous research has shown that students often struggle to develop an
understanding of linear and quadratic relationships. Covariational
reasoning has been identified as a way to support this development. This
study aims to investigate how covariational reasoning supports students
in developing understandings of linear and quadratic relationships
within a projectile motion context. A teaching experiment was conducted
with two middle school students who engaged in a digital task exploring
the relationship between height and time. The analysis characterizes how
the students' covariational reasoning evolved as they interpreted the
changing quantities in the task. The findings suggest that prompts
encouraging students to compare linear and quadratic relationships can
foster more sophisticated forms of covariational reasoning. The
discussion highlights how specific features of the task design,
including the affordances of technology, the emphasis on conceiving
graphs as representations of covarying quantities, and the use of
non-canonical graphing tasks, can support covariational reasoning.
\end{abstract}
{\bfseries \emph Keywords}
\def\sep{\textbullet\ }
covariational reasoning \sep linear relationship \sep projectile
motion \sep 
quadratic relationship

\section{Introduction}\label{introduction}

Linear and quadratic relationships are essential for interpreting
real-world phenomena (Abrams 2023; Stewart, Redlin, and Watson 2016),
yet many studies have shown that students continue to struggle in
understanding these relationships (Van Dooren et al. 2008). For
instance, students often struggle to connect the algebraic forms of
linear and quadratic relationships with their graphical representations
and the real-world situations they model (Leinhardt, Zaslavsky, and
Stein 1990; Wilkie 2019; Wilkie and Ayalon 2018). These challenges
suggest the need for instructional approaches that help students develop
deeper, more connected understandings of how quantities change together,
such as through covariational reasoning.

Covariational reasoning, or reasoning about how two quantities change in
tandem, is fundamental for students' understanding of mathematical
relationships (Thompson and Carlson 2017). Using this reasoning allows
them to interpret functions dynamically, bridging algebraic expressions,
graphs, and contextual situations. Prior studies have shown that
promoting covariational reasoning can help students develop an
understanding of the rate of change (Tallman, Weaver, and Johnson 2024;
Franklin Yu 2024), concepts that underlie both linear and quadratic
relationships. Consequently, fostering covariational reasoning may be a
productive way to address students' challenges in understanding linear
and quadratic relationships.

Therefore, this study employed a teaching experiment to characterize how
students' covariational reasoning supports their development of
understanding of linear and quadratic relationships. We drew on the
covariational reasoning framework (Thompson and Carlson 2017) to analyze
how students reasoned about varying quantities as they engaged with
tasks involving these relationships. This analysis allowed us to trace
how their reasoning evolved and supported conceptual development. The
study offers empirical insights into the role of covariational reasoning
in fostering students' understanding of fundamental functional
relationships in mathematics. The following sections review relevant
literature and theoretical perspectives before describing the methods
and presenting the findings.

\subsection{Literature Review}\label{literature-review}

Previous research has highlighted the potential of covariational
reasoning to support students in developing meaningful understandings of
mathematical relationships. One productive approach for fostering such
reasoning is through the use of real-world phenomena. For example,
Sandoval Jiménez and Sierra (2025) designed a modeling task based on a
Newtonian problem in which students were asked to calculate temperatures
beyond the range measurable by a thermometer. They found that engaging
in this modeling task promoted students' covariational reasoning.
Similarly, F. Yu (2025) and Altindis (2025) incorporated motion-based
contexts into their tasks to foster covariational reasoning. Such
contexts draw on students' intuitive understandings of physical
phenomena, allowing them to focus on coordinating quantities that are
familiar to them (F. Yu 2025). Such real-world contexts can also
generate an intellectual need for students to make sense of mathematical
ideas through covariational reasoning (Harel 2013; Paoletti, Moore, and
Vishnubhotla 2024).

However, students often encounter difficulties when modeling real-world
phenomena, particularly those involving motion contexts. A common
challenge is that students tend to interpret graphs representing
quantities in such phenomena using a spatial coordinate system, even
when this interpretation is inappropriate (Paoletti, Hardison, and Lee
2022). In other words, students often treat the graph as a literal
depiction of the phenomenon, interpreting it as a picture or map of the
physical situation rather than as a representation of how quantities
covary. One promising approach to address this issue is to encourage
students to conceive of a graph both in terms of what is produced (a
trace) and how it is produced (via covarying quantities) (Moore and
Thompson 2015). Another is to employ non-canonical graphing tasks, for
example, those that represent time on the vertical axis and height on
the horizontal axis. These unconventional representations can challenge
students' assumptions and engage them in novel reasoning processes,
including quantitative and covariational reasoning, even when they are
familiar with conventional graph forms (Moore et al. 2014). By
disrupting familiar visual associations, such tasks can promote more
flexible and conceptual understandings of functional relationships.

Previous studies have also leveraged the affordances of technology to
promote students' covariational reasoning in modeling dynamic
situations. For example, Johnson, McClintock, and Gardner (2020) used
technology to display a video animation of a moving car alongside a
sketching panel that allowed students to construct graphs representing
the quantities involved in the situation. This design provided
opportunities for students to engage in covariational reasoning.
Pittalis, Sproesser, and Demosthenous (2025) employed similar
technological affordances to facilitate students in graphically
representing motion scenarios involving covarying quantities. However,
their task design further utilized the affordance of directly linking
the graph to the animation, allowing students to observe how changes in
one representation corresponded to changes in the other. Such uses of
technology illustrate its potential not only to visualize dynamic
relationships but also to create integrated environments where students
can coordinate multiple representations (Ainsworth 1999), thereby
deepening their understanding of how quantities change together.

Taken together, these studies suggest that engaging students with
real-world phenomena, encouraging them to conceive of graphs as
representations of covarying quantities, and introducing non-canonical
graphing tasks can provide meaningful contexts for promoting
covariational reasoning. Moreover, the affordances of technology can
further enhance these learning experiences by enabling students to
dynamically link representations and observe quantitative changes in
real time.

\subsection{Theoretical framework}\label{theoretical-framework}

Covariational reasoning refers to the ability to mentally envision two
or more quantities changing at the same time, maintaining a continuous
awareness of their values and how those values vary together (Saldanha
and Thompson 1998). Thompson and Carlson (2017) describe covariational
reasoning as progressing through several developmental levels,
summarized in Table~\ref{tbl-cov-framework}, ranging from a lack of
coordination between varying quantities to a sophisticated understanding
of two variables varying smoothly and continuously.

\begin{longtable}[]{@{}
  >{\raggedright\arraybackslash}p{(\linewidth - 2\tabcolsep) * \real{0.2500}}
  >{\raggedright\arraybackslash}p{(\linewidth - 2\tabcolsep) * \real{0.7500}}@{}}
\caption{Covariational reasoning framework (Thompson and Carlson
2017)}\label{tbl-cov-framework}\tabularnewline
\toprule\noalign{}
\begin{minipage}[b]{\linewidth}\raggedright
Level
\end{minipage} & \begin{minipage}[b]{\linewidth}\raggedright
Description
\end{minipage} \\
\midrule\noalign{}
\endfirsthead
\toprule\noalign{}
\begin{minipage}[b]{\linewidth}\raggedright
Level
\end{minipage} & \begin{minipage}[b]{\linewidth}\raggedright
Description
\end{minipage} \\
\midrule\noalign{}
\endhead
\bottomrule\noalign{}
\endlastfoot
No coordination & The person has no image of variables varying together.
The person focuses on one or another variable's variation with no
coordination of values. \\
Pre-coordination of values & The person envisions two variables' values
varying, but asynchronously---one variable changes, then the second
variable changes, then the first, and so on. The person does not
anticipate creating pairs of values as multiplicative objects. \\
Gross coordination of values & The person forms a gross image of
quantities' values varying together, such as ``this quantity increases
while that quantity decreases.'' The person does not envision that
individual values of quantities go together. Instead, the person
envisions a loose, non-multiplicative link between the overall changes
in two quantities' values. \\
Coordination of values & The person coordinates the values of one
variable (\(x\)) with values of another variable (\(y\)) with the
anticipation of creating a discrete collection of pairs
\(\left(x, y\right)\). \\
Chunky continuous covariation & The person envisions changes in one
variable's value as happening simultaneously with changes in another
variable's value, and they envision both variables varying with chunky
continuous variation. \\
Smooth continuous covariation & The person envisions increases or
decreases (hereafter, changes) in one quantity's or variable's value
(hereafter, variable) as happening simultaneously with changes in
another variable's value, and the person envisions both variables
varying smoothly and continuously. \\
\end{longtable}

The framework of covariational reasoning presented in
Table~\ref{tbl-cov-framework} does not explicitly include the rate of
change. According to Thompson and Carlson (2017), conceptualizing rate
of change requires covariational reasoning but also involves more
conceptualizations, such as ratio and quotient. Accordingly, attending
to how students quantify a rate of change can offer valuable insights
into their covariational reasoning and inform how instruction might
better support students in reasoning at more advanced levels (Franklin
Yu 2024).

\subsection{Current study}\label{current-study}

The present study aims to explore how covariational reasoning supports
students in developing an understanding of linear and quadratic
relationships in a projectile motion context. Therefore, this study
addresses the following research question: How does covariational
reasoning support middle school students in developing meanings for
linear and quadratic relationships in a projectile motion context?

\section{Methods}\label{methods}

This study employed a teaching experiment (Steffe and Thompson 2000) to
characterize two students' activity as they engaged with the Height-Time
Relationship Task. The task was part of a larger project aimed at
promoting emergent graphical shape thinking, in which students interpret
and construct graphs representing covarying quantities (Kristanto and
Lavicza 2025; Kristanto, Lavicza, and Martinelli 2025). The teaching
experiment was conducted by the first author, who served as the
teacher-researcher (TR) and facilitated the students' engagement with
the task. In this study, we examined how covariational reasoning
supported students in developing meanings for linear and quadratic
relationships. The following sections describe the participants, the
task, and the analytical approach.

\subsection{Participants}\label{participants}

The participants were two female ninth-grade students from a public
middle school in Yogyakarta, Indonesia. TR was not the students' regular
classroom teacher, which helped create an exploratory learning
environment distinct from their usual classroom setting. The students,
referred to by the pseudonyms Fania and Bianca, were selected based on
two criteria. First, they represented diverse academic abilities.
Second, both demonstrated strong verbal communication skills, which were
important for generating rich verbal data during the teaching
experiment. The selection of participants was assisted by the students'
regular mathematics teacher, who identified candidates meeting these
criteria. The participating school was chosen for its familiarity with
using technology, such as laptops or tablets, in classroom instruction.

\subsection{The Height-Time Relationship
Task}\label{the-height-time-relationship-task}

The Height-Time Relationship Task was part of a larger digital task
sequence titled ``How Accurate Is Your Aim?''\footnote{The complete
  digital task sequence can be accessed online at
  \url{https://classroom.amplify.com/activity/69061b958b3deaa21d53e9ef}}
which was designed to provide a game-based learning environment that
supports students' engagement in emergent graphical shape thinking as
they interpret and construct graphs related to projectile motion. The
task sequence was developed in Desmos Classroom (now Amplify Classroom).

The design of the task sequence was informed by suggestions from
previous research. First, it employed a real-world context, projectile
motion, that could serve as a meaningful reference for students as they
engaged in mathematical reasoning (Czocher, Hardison, and Kularajan
2022; Sokolowski 2021). Second, it included non-canonical tasks that
presented quantities in unfamiliar ways. For example, in typical
projectile motion problems, time is represented on the \(x\)-axis and
height on the \(y\)-axis. In this task sequence, the axes were reversed,
with height represented on the \(x\)-axis and time on the \(y\)-axis.
This unconventional representation was intended to engage students in
new reasoning processes, such as quantitative and covariational
reasoning (Moore et al. 2014). Third, the sequence also aimed to promote
students' engagement in emergent graphical shape thinking as they
interpreted and constructed graphs (Moore and Thompson 2015; Paoletti,
Gantt, and Corven 2023). Finally, the design utilized the affordances of
technology to synchronize the animation of the physical phenomenon with
its graphical representation (Ainsworth 1999; Johnson, McClintock, and
Gardner 2020; Pittalis, Sproesser, and Demosthenous 2025). For instance,
when the ball moved upward, the line segment representing its height
became longer, and when the ball descended, the line segment became
shorter. These features were intended to help students connect the
motion they observed with its quantitative representation.

\begin{figure}

\centering{

\pandocbounded{\includegraphics[keepaspectratio]{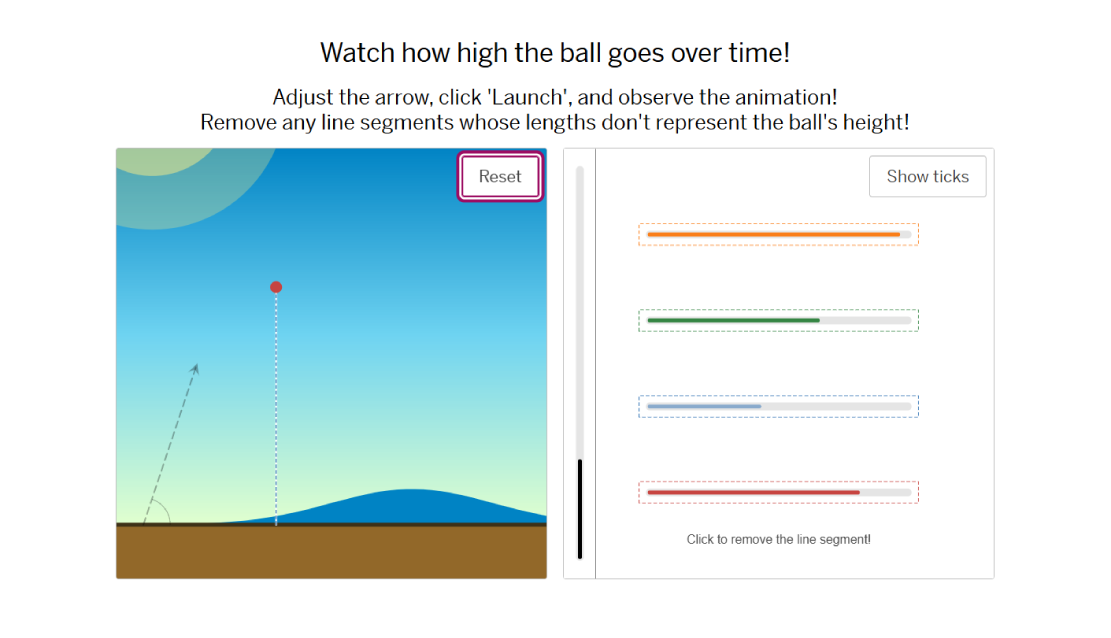}}

}

\caption{\label{fig-screen-1}Screen showing the launch activity where
students select a line segment representing the ball's height over time}

\end{figure}%

Within this sequence, the Height-Time Relationship Task consisted of
three interactive screens that guided students to explore relationships
between height and time during the motion of a launched ball. On the
first screen (Figure~\ref{fig-screen-1}), students could adjust the
length and direction of an arrow to launch a ball. As the ball moved
through its trajectory, they were asked to observe its height over time
and to choose one of four line segments that best represented the ball's
changing height. To support decision-making, students could eliminate
the options one by one until only a single line segment remained.

\begin{figure}

\centering{

\pandocbounded{\includegraphics[keepaspectratio]{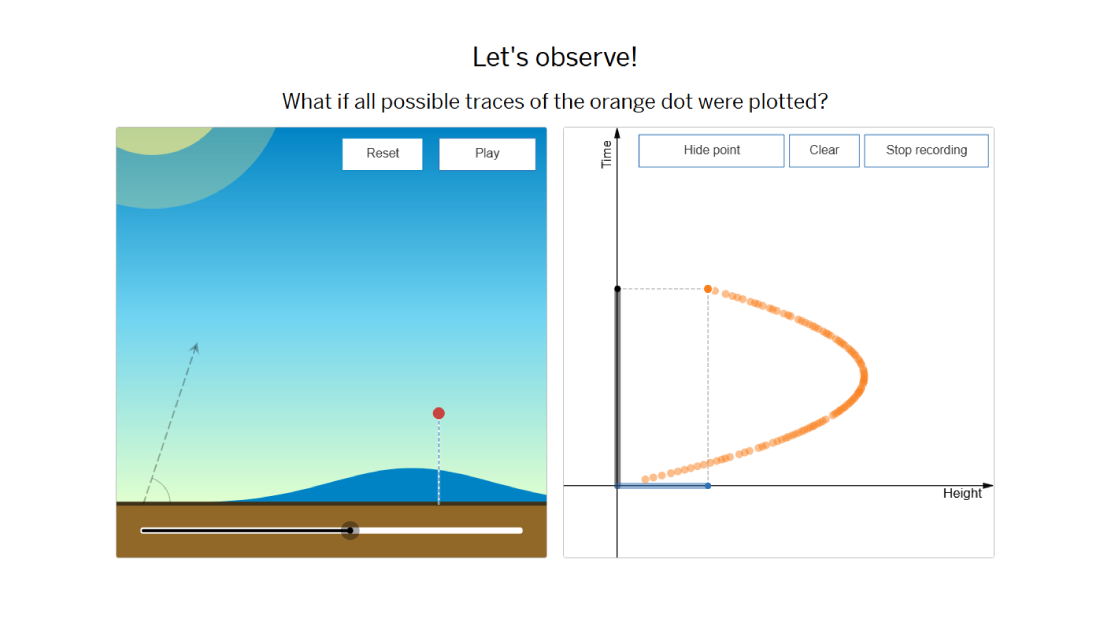}}

}

\caption{\label{fig-screen-2}Screen showing the dynamic relationship
between height and time using moving line segments}

\end{figure}%

On the second screen (Figure~\ref{fig-screen-2}), students again
launched the ball and observed how its height changed over time. In this
activity, the two quantities, height and time, were represented as the
lengths of line segments along the \(x\)- and \(y\)-axes. Students could
represent the relationship between these quantities as a moving point
whose motion was constrained by the two line segments on the axes. They
also had the option to trace the path of the moving point, allowing them
to visualize the trajectory it followed.

\begin{figure}

\centering{

\pandocbounded{\includegraphics[keepaspectratio]{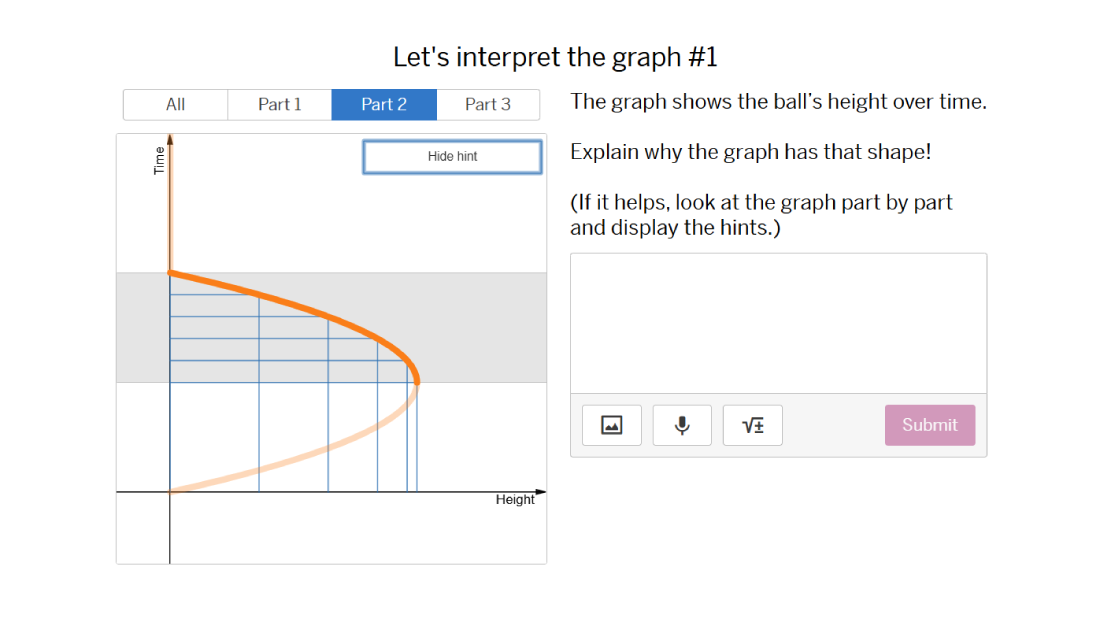}}

}

\caption{\label{fig-screen-3}Graph interpretation screen where students
analyze the ball's height-time graph}

\end{figure}%

The third screen (Figure~\ref{fig-screen-3}) displayed a graph showing
the height of the ball as a function of time and prompted students to
interpret the shape of the graph. Students could analyze the graph in
sections and were offered hints consisting of several equally spaced
points on the \(y\)-axis, along with their corresponding points on the
\(x\)-axis. These hints supported students in connecting graphical
features to the quantitative relationships represented.

\subsection{Analysis}\label{analysis}

During data collection, the two students sat side by side and worked
collaboratively on a single laptop to complete the task. TR sat in front
of them to facilitate the teaching experiment. While the students
worked, their laptop screen was recorded using screen recording
software, and their facial expressions were captured using the built-in
webcam. A separate camera positioned behind them recorded their gestures
and audio. All recordings were synchronized into a single composite
video to support detailed analysis. Based on this video, we selectively
transcribed the dialogue between TR and the students, along with the
students' gestures, focusing on their activity on the third screen of
the task. In particular, we focused on coverbal gestures, or gestures
produced in coordination with speech, as these often convey meanings
that complement verbal expressions and reflect students' emerging
reasoning processes (Kendon 2000; Kong, Law, and Chak 2017; Paneth et
al. 2024).

In analyzing the data, we first identified instances in the students'
activity that provided evidence of their covariational reasoning as they
engaged with the third screen. This screen was chosen because it was
where the students developed their understanding of linear and quadratic
relationships through graph interpretation. Second, we coded these
instances using Thompson and Carlson's (2017) covariational reasoning
framework. To strengthen the reliability of our analysis, we maintained
a clear evidentiary trail that linked the data and our interpretations
directly to the research questions. Finally, we connected the students'
covariational reasoning with how they constructed understandings for
linear and quadratic relationships.

\section{Findings and Discussion}\label{findings-and-discussion}

This section presents the activity of Fania and Bianca as they engaged
with the third screen of the Height-Time Relationship Task
(Figure~\ref{fig-screen-3}). The findings are reported chronologically
to trace the shift in their reasoning. First, we describe how the
students interpreted what the graph was composed of without yet
connecting it to the covarying quantities it represented. Second, we
examine their emerging understandings of the two covarying quantities
represented in the graph. Finally, we present how their covariational
reasoning developed as they compared the graph with a linear function
graph.

\subsection{Interpreting the Structure of the
Graph}\label{interpreting-the-structure-of-the-graph}

When Fania and Bianca opened the third screen, TR asked them to describe
what the displayed graph represented. Responding to this prompt, Fania
said:

\begin{quote}
Because it's like a collection of recorded data. Uh, what is it\ldots{}
possibilities. A trace, like a record of all the possible points where
the points could be.
\end{quote}

We interpreted Fania's understanding as being influenced by her
experience on the previous screen (Figure~\ref{fig-screen-2}). On that
screen, she had observed how a graph representing the ball's height over
time was formed as a trace of a moving point whose motion was determined
by the quantities it represented, namely height and time. Based on her
statement, we infer that Fania conceived of the graph as traces of a
point whose position was determined by the two varying quantities, the
ball's height and time. This interpretation of a graph as a trace of
dynamic quantities is consistent with the perspectives suggested by
Moore and Thompson (2015) and Paoletti, Gantt, and Corven (2023).

TR then asked what shape the graph had. Fania responded:

\begin{quote}
Like a half circle, curved, kind of oval {[}makes an oval gesture with
her right thumb and index finger{]}, and then it's straight {[}uses her
left hand to make a vertical line gesture{]}.
\end{quote}

Bianca agreed with Fania's interpretation. Their description of the
graph's shape was notable for two reasons. First, neither of them
referred to it as a parabola, which suggests that the graph appeared
novel to them. Second, their use of the term half oval to describe its
form indicates that they drew on a familiar visual shape, likely a
horizontally oriented half oval, to describe what they observed.

\subsection{Constructing Meanings for Covarying
Quantities}\label{constructing-meanings-for-covarying-quantities}

TR then invited Fania and Bianca to focus more closely on interpreting
the shape of the graph. Specifically, TR asked why the graph appeared as
a half oval at first and was then followed by a vertical line. This
question initiated the discussion presented in
Table~\ref{tbl-excerpt-1}.

\begin{longtable}[]{@{}
  >{\raggedright\arraybackslash}p{(\linewidth - 2\tabcolsep) * \real{0.2500}}
  >{\raggedright\arraybackslash}p{(\linewidth - 2\tabcolsep) * \real{0.7500}}@{}}
\caption{Excerpt of students' discussion when interpreting the shape of
the graph}\label{tbl-excerpt-1}\tabularnewline
\toprule\noalign{}
\begin{minipage}[b]{\linewidth}\raggedright
Speaker
\end{minipage} & \begin{minipage}[b]{\linewidth}\raggedright
Transcript
\end{minipage} \\
\midrule\noalign{}
\endfirsthead
\toprule\noalign{}
\begin{minipage}[b]{\linewidth}\raggedright
Speaker
\end{minipage} & \begin{minipage}[b]{\linewidth}\raggedright
Transcript
\end{minipage} \\
\midrule\noalign{}
\endhead
\bottomrule\noalign{}
\endlastfoot
Fania & Mmm, maybe it's also affected by height. Because at the
beginning, it starts from below {[}places right hand low{]}, then it's
thrown {[}makes a gesture of throwing a ball upward{]}. Since it's
thrown, the height on the graph moves to the right. At a certain point,
mmm, it comes back down because it's pulled by gravity {[}uses hand to
trace the motion of a ball going up and down like a parabola{]}. So it
comes back, and the graph moves to the upper left {[}uses right hand to
make a diagonal gesture upward to the left{]}. That means the ball is
going down {[}mimics a falling motion with her hand{]}. Then, when it
reaches the ground, there's no more change in the ball's height, but
time keeps moving, so the graph becomes a straight line. \\
TR & Bianca, anything to add? \\
Bianca & {[}Shakes her head{]} No, it's the same. \\
\end{longtable}

Based on the discussion in Table~\ref{tbl-excerpt-1}, Fania was able to
describe how the quantities of time and the ball's height changed
together. In doing so, she coordinated the physical situation, the
motion of the ball over time, with the graph that represented it. For
instance, she associated the upward-left segment of the graph with the
moment when the ball was descending. This interpretation provides strong
evidence that she was able to make meaningful sense of the graph and
engage in covariational reasoning. Rather than construing the coordinate
system spatially, which is a common tendency among students who
interpret graphs as literal depictions of physical trajectories (Clement
1989; Hadjidemetriou and Williams 2002; Paoletti et al. 2022), she
conceptualized it quantitatively, with height represented on the
\(x\)-axis and time on the \(y\)-axis. Within this understanding, she
appropriately interpreted the upward-left segment of the graph as
representing a decrease in height as time progressed, corresponding to
the ball's downward motion. However, there was no evidence that she
constructed a multiplicative object that explicitly paired time and
height values. Therefore, we infer that she was engaging in
covariational reasoning, particularly at the stage of gross coordination
of values.

\subsection{Shifts in Covariational Reasoning through Comparison with a
Linear
Graph}\label{shifts-in-covariational-reasoning-through-comparison-with-a-linear-graph}

\begin{figure}

\centering{

\pandocbounded{\includegraphics[keepaspectratio]{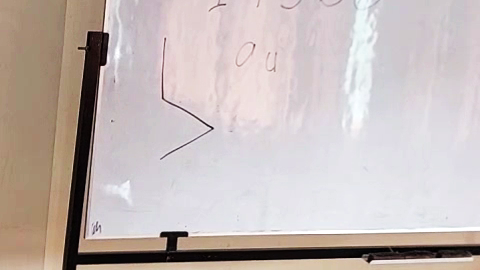}}

}

\caption{\label{fig-tr-graph}The graph that was drawn by TR on the
whiteboard}

\end{figure}

To promote more advanced covariational reasoning, TR invited Fania and
Bianca to compare the graph on the third screen with a graph that the TR
drew on the whiteboard. On the whiteboard, TR sketched a graph that
followed the same overall pattern previously described by Fania: first
extending upward to the right, then upward to the left, and finally
continuing with a vertical line upward. However, unlike the digital
graph, all parts of TR's graph followed linear patterns
(Figure~\ref{fig-tr-graph}).

TR then asked why the graph on the third screen appeared curved rather
than composed of straight segments, as in Figure~\ref{fig-tr-graph}.
This question prompted Fania and Bianca to interpret TR's graph and led
to the discussion presented in Table~\ref{tbl-excerpt-2}.

\begin{longtable}[]{@{}
  >{\raggedright\arraybackslash}p{(\linewidth - 2\tabcolsep) * \real{0.2500}}
  >{\raggedright\arraybackslash}p{(\linewidth - 2\tabcolsep) * \real{0.7500}}@{}}
\caption{Discussion comparing the curved and linear
graphs}\label{tbl-excerpt-2}\tabularnewline
\toprule\noalign{}
\begin{minipage}[b]{\linewidth}\raggedright
Speaker
\end{minipage} & \begin{minipage}[b]{\linewidth}\raggedright
Transcript
\end{minipage} \\
\midrule\noalign{}
\endfirsthead
\toprule\noalign{}
\begin{minipage}[b]{\linewidth}\raggedright
Speaker
\end{minipage} & \begin{minipage}[b]{\linewidth}\raggedright
Transcript
\end{minipage} \\
\midrule\noalign{}
\endhead
\bottomrule\noalign{}
\endlastfoot
Fania & If it's diagonal, if it's straight like that {[}pointing to the
TR's graph on the whiteboard{]}, that means the speed is constant. \\
Bianca & Yes. \\
Fania & Every second it-- \\
Bianca & Every second it-- \\
Fania & Same. \\
Bianca & Same. \\
\end{longtable}

Based on Table~\ref{tbl-excerpt-2}, Fania and Bianca identified the key
characteristic of the graph drawn by TR, namely that it represented a
linear function. They interpreted that such a graph depicts two
quantities whose average change is constant. In this reasoning, they
were not only engaged in covariational reasoning but also made use of
the ratio between the change in one quantity, the ball's height, and the
change in the other quantity, time. Specifically, they used ``every
second'' as the unit of time. Therefore, we infer that they were engaged
in covariational reasoning at least at the stage of chunky continuous
covariation.

After that, TR asked Fania and Bianca to interpret the graph on the
third screen again, but this time part by part, using the hints
provided. The following excerpt shows Fania's interpretation of the
first part of the graph, which represents the ball's height over time as
it moves upward.

\begin{quote}
Oh, that means within the same unit of time, the height isn't
necessarily the same. If the graph were a straight diagonal line like
that {[}pointing to the graph on the whiteboard{]}, then within the same
time interval, the height would always be the same, so the differences
between intervals would be equal. But because this graph curves, that
means for the same unit of time, the height varies. So, in the first
second, for example, it's still far apart because it's just been thrown
{[}making a throwing gesture from bottom to top{]}, so the height
increases quickly. In the second second, it gets slower {[}gesturing
smaller intervals with her right thumb and index finger{]}. So the graph
becomes more compact, see {[}gesturing again to show smaller intervals
with her left hand{]}. In the third second, it's even smaller because it
slows down again {[}repeating the gesture with her right hand{]}. In the
fourth second, smaller again {[}same gesture{]}. Until finally it stops
at the top {[}pointing upward with her right hand{]}, then goes down
again {[}mimicking the downward motion of the ball{]}.
\end{quote}

TR then invited Bianca to add to or restate Fania's explanation in her
own words. Bianca responded as follows:

\begin{quote}
So, if the diagonal is a bit curved, that means each second has a
different height, not the same. If it's a straight line like this
{[}moving her right hand upward diagonally to form a straight line{]},
that means every second the height is the same, like it goes straight up
{[}repeating the same gesture{]}. And, what's it called, when the
distance between them gets smaller {[}gesturing smaller intervals with
her right thumb and index finger{]}, that means it's about to go down.
\end{quote}

Based on Fania and Bianca's explanations, they contrasted the graph on
the third screen with the linear graph drawn on the whiteboard.
Specifically, they compared how one quantity changed while the other
varied at a constant rate. They interpreted that the ball's height
changed by progressively smaller amounts as time increased uniformly.
Fania referred to ``the first second,'' ``the second second,'' ``the
third second,'' and ``the fourth second'' to indicate that she was
considering equal units of time, even though the intervals shown by the
hints on the third screen did not necessarily represent one second.
Thus, we infer that both students engaged in covariational reasoning,
and that their reasoning showed a shift toward a more sophisticated
coordination of the two varying quantities. Whereas earlier they
demonstrated reasoning at the stage of gross coordination of values,
they now exhibited reasoning characteristic of the chunky continuous
covariation stage.

Fania and Bianca applied similar reasoning when interpreting the second
part of the graph, which represented the ball's descent. Their
explanations reflected covariational reasoning at the stage of chunky
continuous covariation. TR then invited one of them to interpret the
final segment of the graph. Bianca responded as follows:

\begin{quote}
This means that the ball is no longer moving and has stopped, but time
keeps passing. {[}She moved her right hand upward to illustrate the
increase in time along the y-axis.{]}
\end{quote}

Bianca's explanation indicates her understanding of the vertical line
segment on the graph. She interpreted the vertical segment as
representing a constant height while time continued to progress. In
doing so, she no longer referred to discrete time units to explain the
ball's motion but instead expressed that the height remained unchanged
as time flowed continuously. We therefore infer that she engaged in
covariational reasoning at the smooth continuous covariation stage,
demonstrating a more refined coordination of continuously varying
quantities.

\subsection{Discussion}\label{discussion}

We have shown how Fania and Bianca's covariational reasoning developed
as they interpreted the relationship between height and time in the
context of projectile motion. Our findings indicate that prompts
encouraging students to compare linear and quadratic relationships can
foster more sophisticated forms of covariational reasoning. Moreover,
such reasoning supported the students in constructing meaning for a
linear relationship as one in which the change in one quantity remains
constant when the change in the other quantity is constant. In other
words, they understood a linear relationship as one characterized by a
constant rate of change.

Building on this understanding, the students developed meanings for the
relationship between height and time in projectile motion: that the
change in height becomes smaller as the ball rises and larger as it
falls. Although this understanding alone does not fully characterize the
quadratic relationship that models projectile motion, it may serve as a
productive entry point, or in Harel's (2013) terms, an intellectual
need, for exploring additional characteristics of quadratic
relationships. Thus, our study contributes to the growing body of
research that employs covariational reasoning as a lens for supporting
students in making sense of different types of mathematical
relationships (Hohensee et al. 2024; Paoletti and Vishnubhotla 2022;
Sokolowski 2021; Wilkie 2020).

In our findings, Fania and Bianca engaged in covariational reasoning
both when imagining the motion of the ball (whether rising, falling, or
stationary) and when interpreting the graph. We highlight several
important points regarding this. First, the opportunity to observe the
physical situation helped them construct the involved quantities and the
relationships between them. In our study, this was supported by the
technological affordances that allowed the situation to be displayed
dynamically while simultaneously showing the corresponding quantitative
representations. This aligns with prior studies demonstrating the role
of technology in promoting covariational reasoning (Johnson, McClintock,
and Gardner 2020; Martínez-Sierra and Villadiego Franco 2025; Pittalis,
Sproesser, and Demosthenous 2025). Second, the students' understanding
of the graph as a representation of covarying quantities, consistent
with the perspective suggested by Moore and Thompson (2015), supported
their engagement in covariational reasoning. This finding resonates with
previous research highlighting the importance of helping students view
graphs as representations of covarying quantities rather than as static
shapes (Moore 2021; Paoletti, Hardison, and Lee 2022; Paoletti et al.
2024). Finally, our study also shows that the use of non-canonical
graphing tasks can be effectively implemented with middle school
students, providing them with opportunities to engage in novel and
productive forms of mathematical reasoning (Moore et al. 2014; Moore et
al. 2019). In addition, the use of such tasks in our study provided
strong evidence for assessing whether students engaged in covariational
reasoning when interpreting graphs. The non-canonical design enabled us
to distinguish whether students were reasoning within a spatial
coordinate system or constructing a quantitative coordinate system as
they analyzed the projectile motion context.

Future research could extend this work by exploring students'
covariational reasoning in analyzing other representations of projectile
motion, such as horizontal distance versus time. In doing so, students'
understanding of height-time and horizontal distance-time relationships
may support their reasoning about parametric representations of
projectile motion. Another promising direction is to further investigate
how students develop an understanding of quadratic relationships as
involving a constant rate of change of the rate of change.

This study has several limitations. The analysis focused only on the
students' activity on the third screen, so it does not capture how their
reasoning may have been influenced by earlier screens of the same task.
However, this focus still provided meaningful insights into how
covariational reasoning supports students in developing an understanding
of linear and quadratic relationships. In addition, the study involved
only two students, which limits the generalizability of the findings, as
other students may develop different ways of understanding these
relationships.

\section{Conclusion}\label{conclusion}

This study has illustrated a developmental shift in how middle school
students engage in covariational reasoning when interpreting
relationships between height and time in a projectile motion context. By
comparing linear and quadratic relationships, students developed an
understanding of constant and varying rates of change, which supported
their construction of meaningful connections between quantities. The
findings highlight the potential of technology-enhanced, non-canonical
graphing tasks for fostering students' covariational reasoning in
dynamic situations.

\section{Acknowledgements}\label{acknowledgements}

This study is part of the first author's doctoral research, which is
supported by an Ernst Mach Grant funded by OeAD-GmbH in cooperation with
ASEA-UNINET.

\section{References}\label{references}

\phantomsection\label{refs}

\end{document}